\documentclass{PoS}

\title{Microhalos and Dark Matter Detection}

\ShortTitle{Microhalos and Dark Matter Detection}

\author{\speaker{Aurel Schneider}$^a$, Lawrence M. Krauss$^b$ and Ben Moore$^a$\\
\llap{$^a$} Institute for Theoretical Physics, University of Zurich, Zurich, Switzerland.\\
\llap{$^b$} School of Earth and Space Exploration and Department of Physics, Arizona State University, PO Box 871404, Tempe, AZ 85287.\\
E-mail: \email{aurel@physik.uzh.ch}}

\abstract{Cosmological structure formation predicts that our galactic halo contains an enormous hierarchy of substructures and streams, the remnants of the merging hierarchy that began with tiny Earth mass microhalos. If these structures persist until the present time, they could influence dramatically the detection signatures of weakly interacting elementary particle dark matter (WIMP). Using numerical simulations that follow the tidal disruption within the Galactic potential and heating from stellar encounters, we find that neither microhalos nor streams have significant impact on direct detection, implying that dark matter constraints derived using simple smooth halo models are relatively robust. We also find that many dense central cusps survive, yielding a small enhancement in the signal for indirect detection experiments.}

\FullConference{Identification of Dark Matter 2010-IDM2010\\
		July 26-30, 2010\\
		Montpellier France}

\begin{document}

\section{Introduction}

Structure formation in a $\Lambda$CDM universe happens in a bottom-up process via hierarchical clustering and merging of small density perturbations. The mass of the smallest and most abundant structures in the universe is modulated by the free streaming velocity of the dark matter particle. For a 100 GeV neutralino these structures, called microhalos, have a mass comparable to the Earth and a half mass radius of $10^{-2}$ pc. \cite{Berezinsky2003,Koushiappas2009}. Since such small objects already formed at redshift $80-30$, they are very dense and therefore the most probable dark matter structures to survive within our galaxy until the present day \cite{Green2004}.

The dominant disruption processes acting on substructures are the tidal forces of the galactic potential and the interaction with stars in the disk. From N-body simulations of galaxy formation we know that these effects are able to completely disrupt most large substructures inside a galactic radius of 30 kpc \cite{dOnghia2010}. Microhalos are more difficult to disrupt and it is possible that a large number of them survive until today within the solar radius. This could then dramatically alter the phase-space distribution of CDM particles and therefore affect dark matter detection experiments.

In order to address the question of the survival of microhalos we will carry out simulations that follow the evolution of these objects and which include the dominant disruption processes in the galaxy. We also calculate the survival statistics of microhaloes using realistic orbital distributions within the disk, allowing us to follow the dynamical structure of the dark matter streams and thus to estimate the fine grained phase space distribution function of WIMPs on scales relevant to dark matter detection experiments.

The work presented here is mainly a summary of the paper \cite{Schneider2010} and readers interested in more details are refered there.

\section{Simulation of the disruption processes}

Orbiting microhalos are mainly affected by tidal effects of the galactic potential and gravitational interactions with single stars during the orbital crossing of the disk. Unfortunatley we cannot simulate both effects simultaneously as this would require setting up a self-consistent disk with billions of stars and dark matter particles. Instead we look seperately at the disk interaction and the tidal disruption and we then estimate the combined effects.

\vspace{0.5cm}
\noindent
\textbf{Disk interaction:} As a substructure halo crosses through a stellar field, high-speed interactions with single stars will heat up the halo distribution, causing it to increase its velocity dispersion and hence its scale size will grow \cite{Goerdt2006,Zhao2007}. We simulated the effect of disk crossing with a microhalo crossing a periodic box of stars with a velocity of $200$ km/s. The stars in the box are randomly distributed with the density $\rho=0.04$M$_{\odot} $pc$^{-3}$ and the velocity dispersion $\sigma = 50$ km/s. This constellation corresponds to the stellar field in the disk at the solar radius \cite{B&T}. For simplicity all the stars have the average mass of $0.7$ M$_{\odot}$. The mass and the density profile of the microhalo corresponds to the result obtained by Diemand et. al. \cite{Diemand2005}.

\begin{figure}[ht]
\begin{minipage}{7.8cm}
\includegraphics[scale=.33]{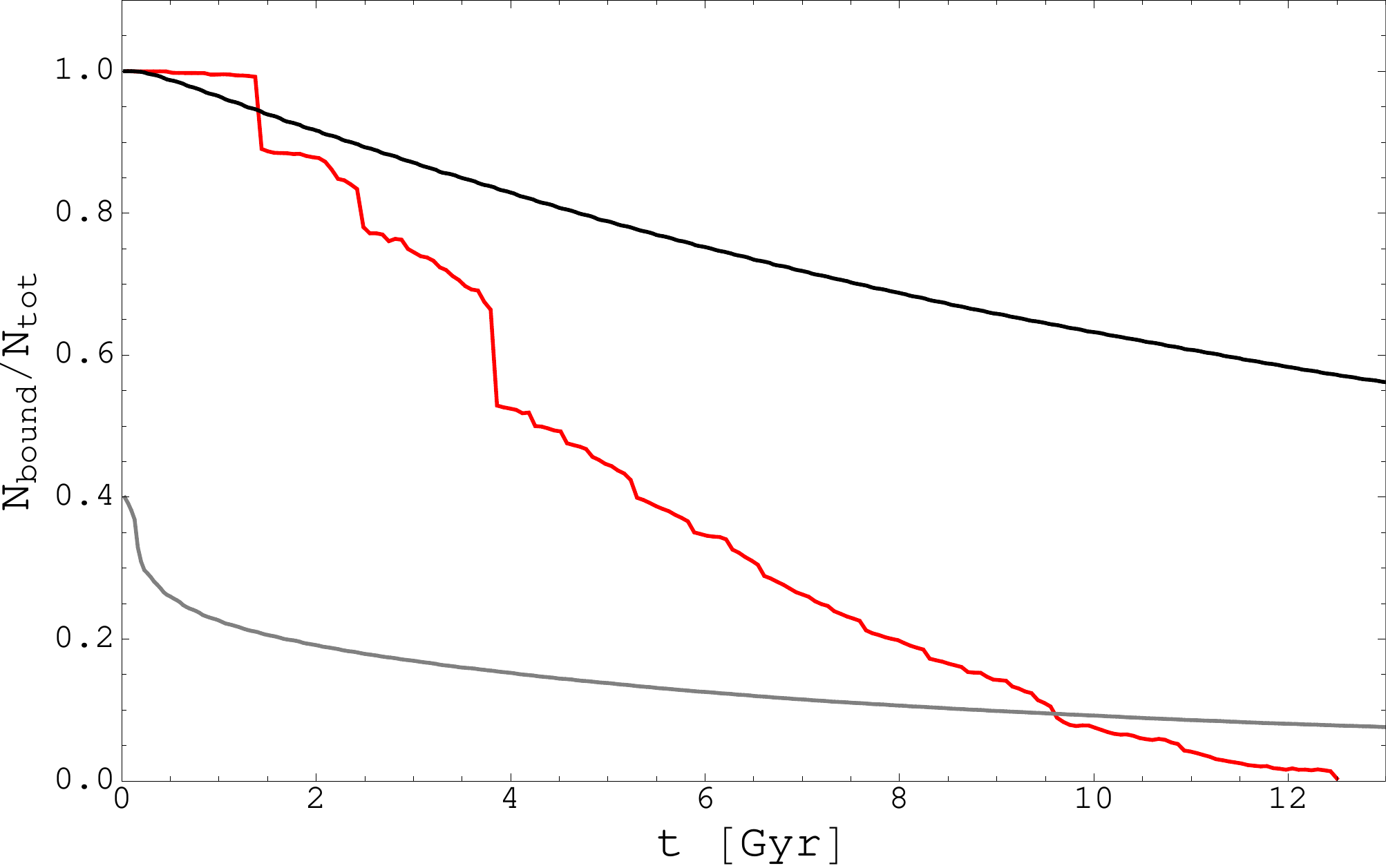}
\end{minipage}
\begin{minipage}{7.8cm}
\includegraphics[scale=.36]{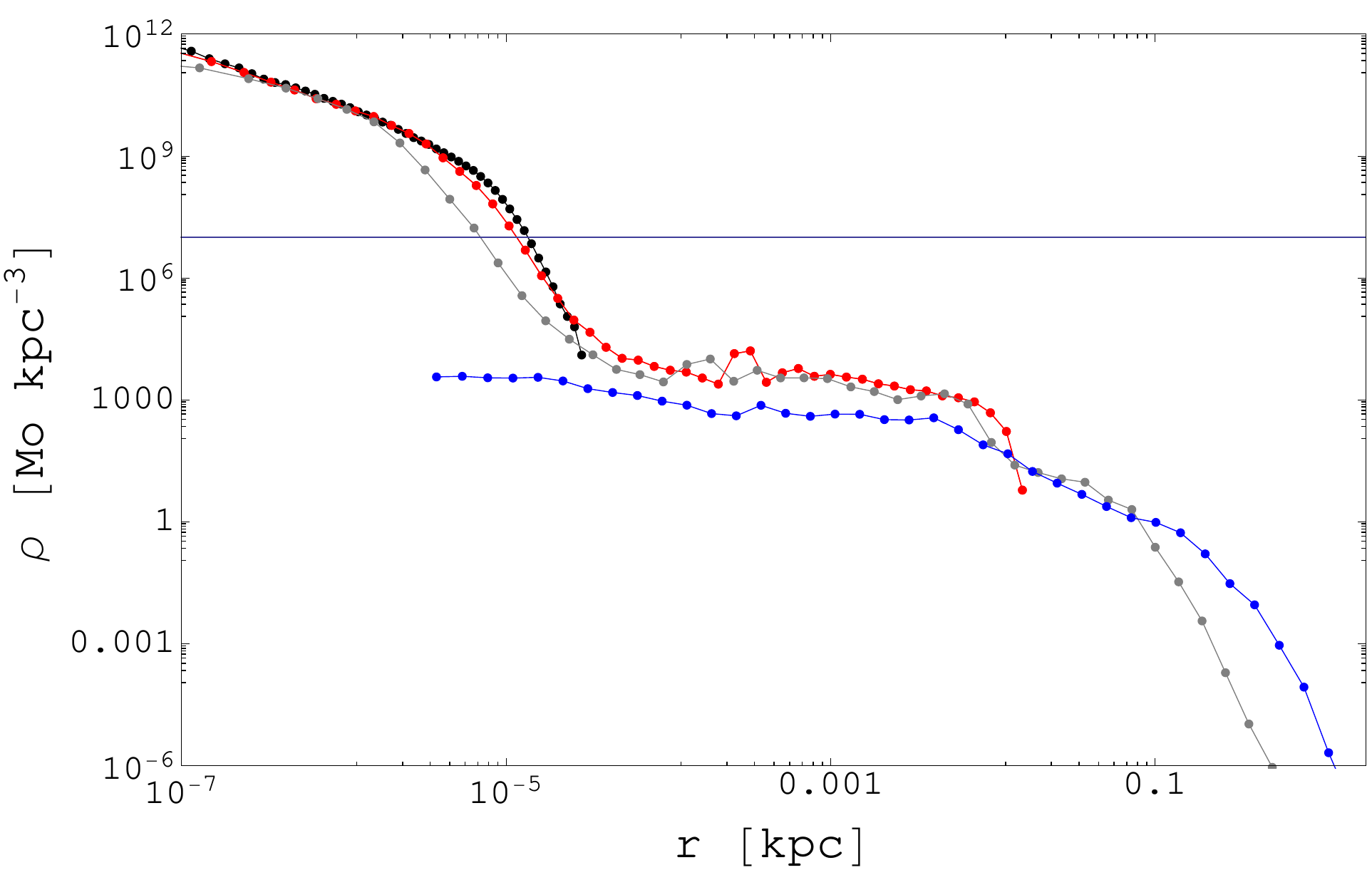}
\end{minipage}
\caption{\textit{Ratio between bound and total mass of a microhalo crossing a stellar field (red) and orbiting in a Milky Way potential after having spent 0 Myr (black) respectively 80 Myr (grey) in the stellar field.}}\label{disruption_plot}
\caption{\textit{Stream densities of microhalos after an orbital time of ten Gyr. Before orbiting the microhalos have spent 0 Myr (red), 80 Myr (grey) and 160 Myr (blue) in the stellar field. The black dots represent the density profile of a completely undisrupted microhalo. The straight blue line corresponds to the average dark matter density at the radius of the sun.}}\label{stream_density}
\end{figure}

In our simulations we find that $50\%$ of the microhalo mass is unbound after 80 Myr of box-crossing (which corresponds to about 40 perpendicular disk passages). After 160 Myr (80 disk passages) even the central core starts to disappear and more than $90\%$ of the microhalo is completely disrupted (see pictures in Table \ref{mh_crossing_den}). At latest after 200 Myr (100 disk passsages) no bound structure is left (Fig. \ref{disruption_plot}).
\begin{table}[ht]
\centering
\begin{tabular}{cccc}
\includegraphics[scale=0.12]{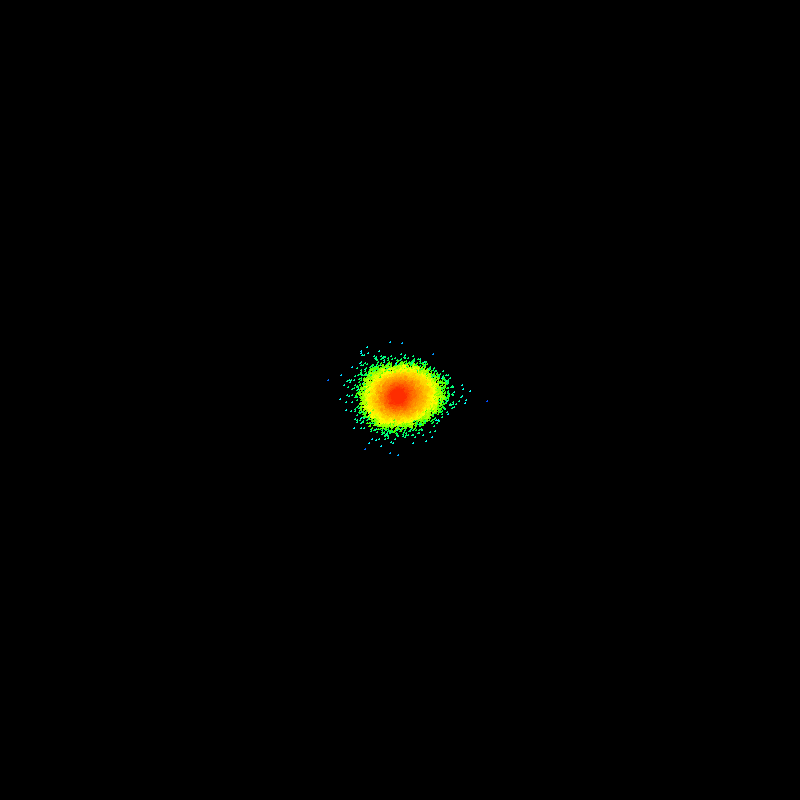}&\includegraphics[scale=0.12]{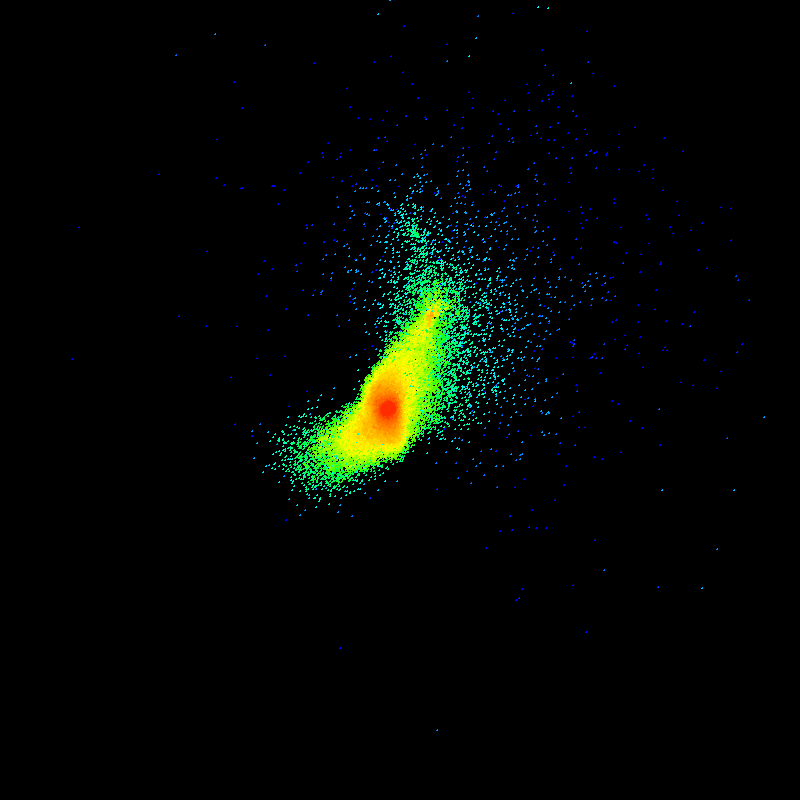}&\includegraphics[scale=0.12]{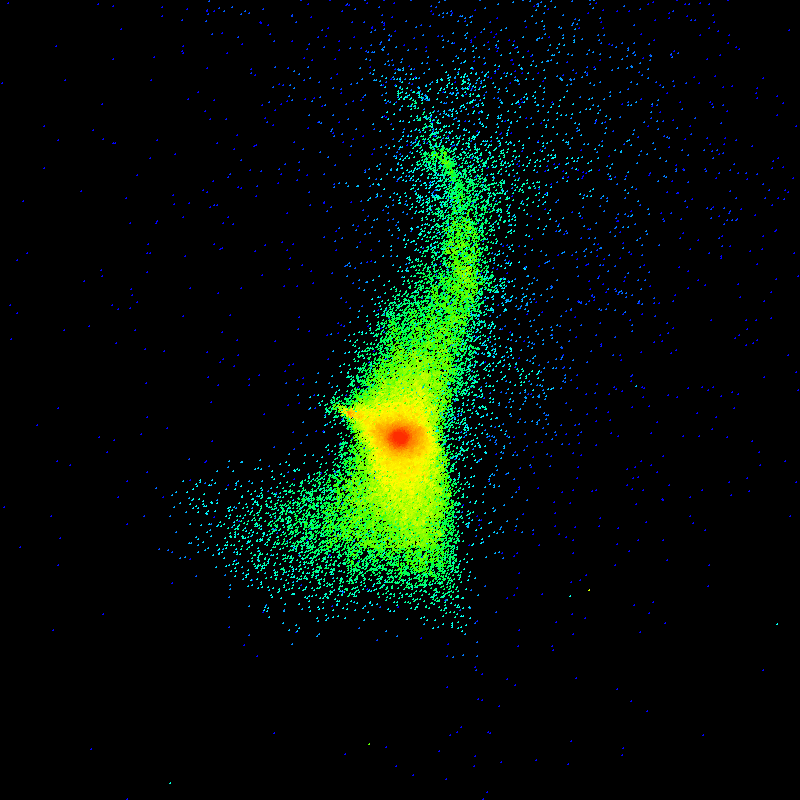}&\includegraphics[scale=0.12]{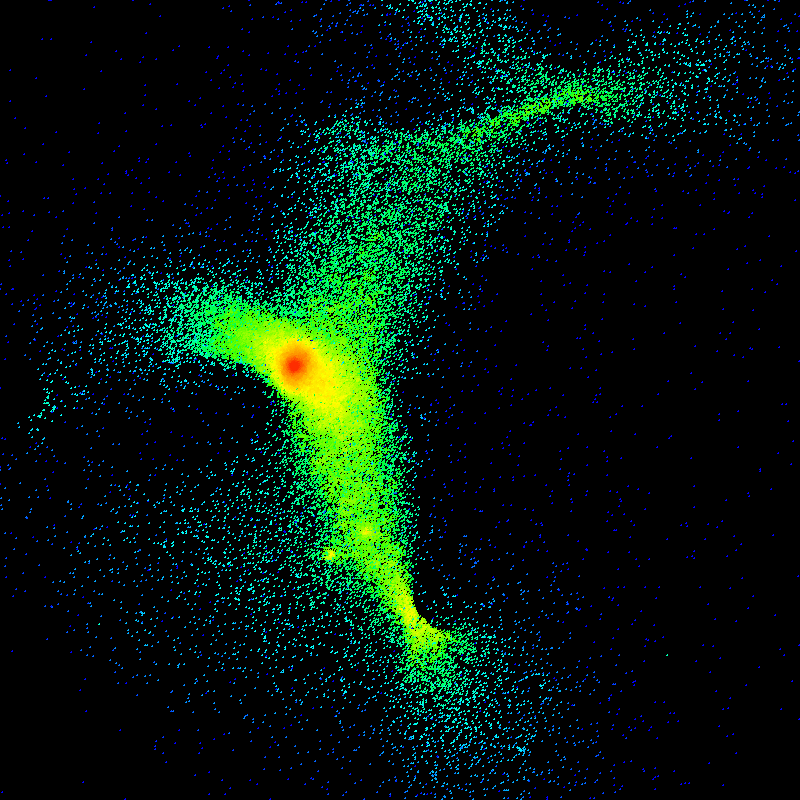}\\
\newline
\includegraphics[scale=0.12]{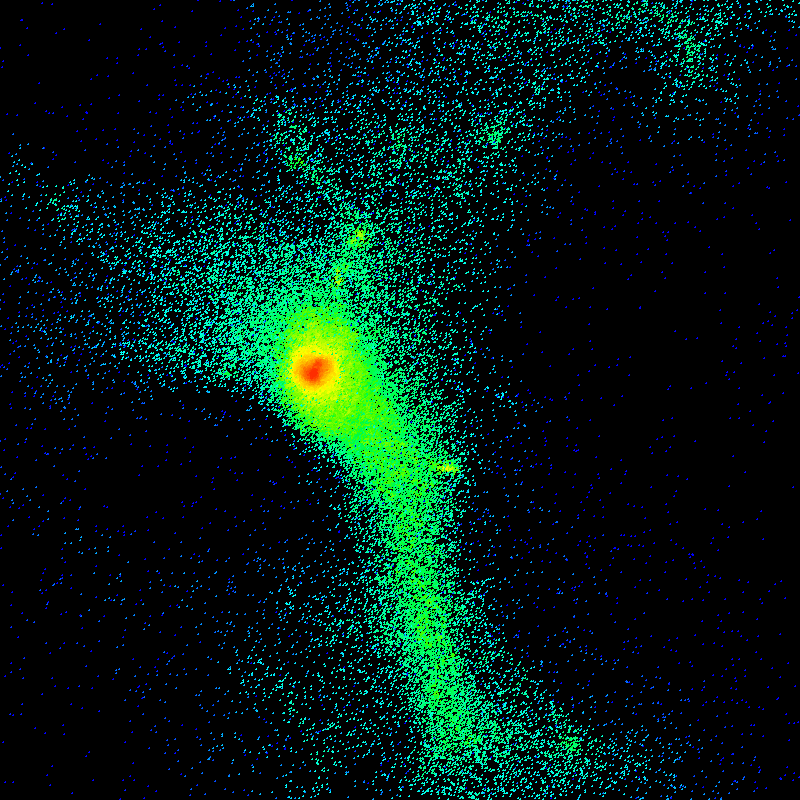}&\includegraphics[scale=0.12]{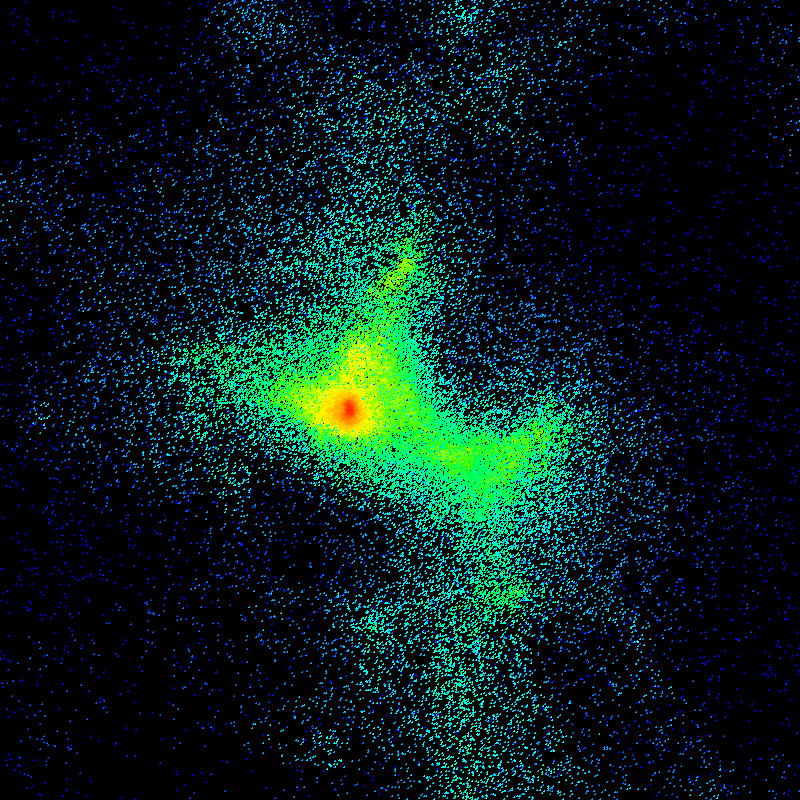}&\includegraphics[scale=0.12]{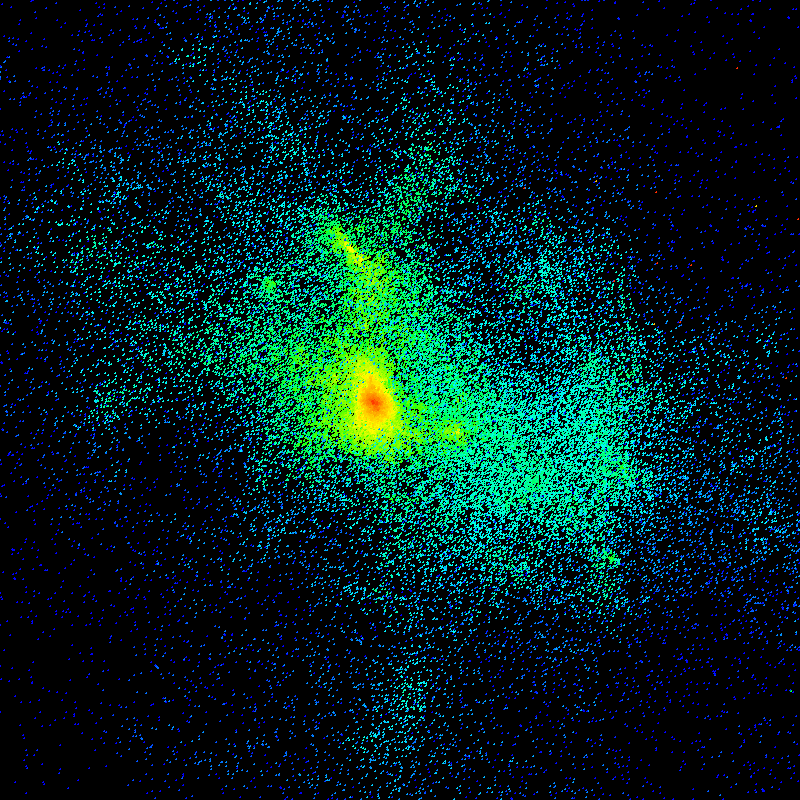}&\includegraphics[scale=0.12]{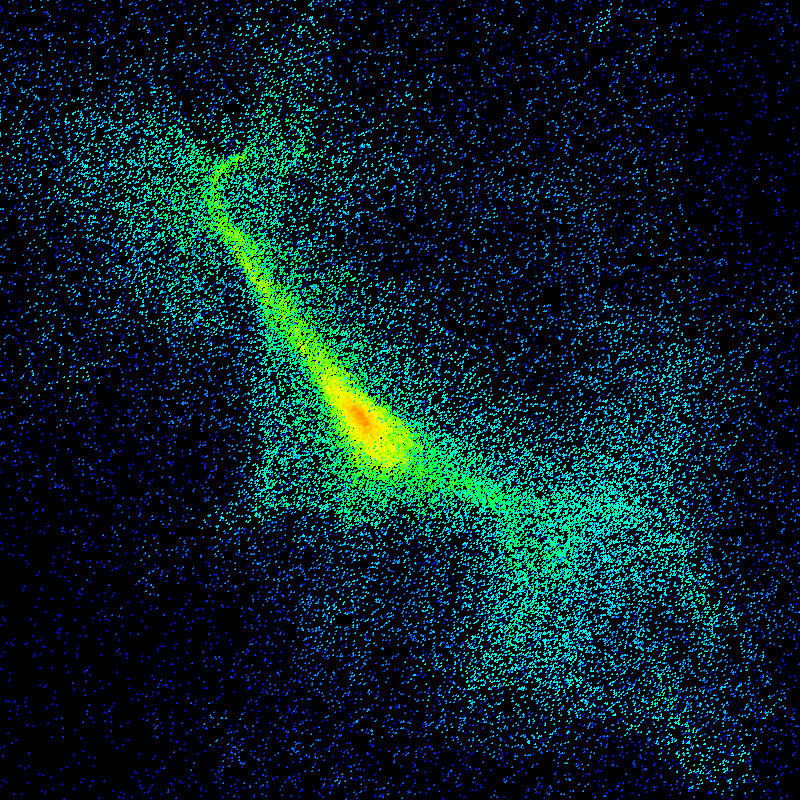}\\
\end{tabular}
\caption{\textit{Microhalo density map at $t=$ 20, 40, 60, 80, 100, 120, 140, 160 Myr (from the upper left to the lower right). The boxlength of the images is 0.38 pc.}}
\label{mh_crossing_den}
\end{table}

An orbital analysis of dark matter particles in an realistic model of our galaxy gives us an average number of disk crossings of $\overline{c}=80$ with a standard deviation $\sigma_{c} =43$. The average crossing radius is (not surprisingly) $\overline{R}=8$ kpc with $\sigma_{R} =4$ kpc. The spread of disk crossing events for different particles follows a Maxwell-Boltzmann distribution.

We use this disk crossing distribution combined with the rate of mass loss determined from our numerical study to calculate the survival statistics of microhalos in the vicinity of the sun. Since the timescale for complete disruption in our simulation is equivalent to the average time a microhalo spends in the stellar disk, we conclude that the average microhalo in the vicinity of the sun is just about to be entirely destroyed at the present time (see also \cite{Green2007}). At most five percent of its initial mass is still in a bound core. However the spread in the number of disc crossings is relatively wide and a significant fraction of microhalos should still have surviving cores. Mass loss is nevertheless important: microhalos maintaining more than $50\%$ of their initial mass should be rare. Figure \ref{disruption_plot} illustrates the mass loss, where the red curve shows the disruption of a typical microhalo with 80 disk crossings in 10 Gyr at the radius of the sun.

\vspace{0.5cm}
\noindent
\textbf{Tidal disruption:} The second main disruption process happens due to tidal forces induced by the galactic potential. While orbiting the galaxy, the microhalo gets truncated and the escaping particles will be assembled in elongated leading and trailing tidal streams. The detailed impact of tidal streaming depends on the orbit of the microhalo and on the shape of the host potential. In our simulations we use a realistical potential of a milky-way galaxy, set up with the galactICs code of Widrow and Dubinsky \cite{Widrow2005}, and we choose roughly spherical orbits with a radius of $7.9$ kpc from the galactic center.

The simulations are performed for three different cases: an initially completely undisturbed microhalo, a microhalo that first crossed the stellar field for 80 Myr and has lost about 60 percent of its mass, and a completely disrupted microhalo that spent more than 160 Myr in the stellar field. In that way we can estimate the combined behavior of the stellar disruption in the disk and the tidal disruption.

Orbiting in the galactic potential significantly reduces the mass of the microhalo (see black and grey lines in Fig \ref{disruption_plot}). However, the rate of tidal mass loss decreases as the tidal radius is steadily reduced. The central cusp of each dark matter microhalo has a very deep potential, as a consequence there is always a bound core remaining, even for a microhalo that has been heated in the stellar field before orbiting.

\vspace{0.5cm}
\noindent
Comparing the curves in Fig \ref{disruption_plot} leads to the conclusion that disk crossing is the dominant disruption process and the only one that can lead to complete distruction of the microhalo. The step-like decrease of the curve is an indication of very close encounters that play a mayor role in the disruption process. Tidal stripping on the other hand can also significantly reduce the mass but it never completely distroys the microhalo because of its tightly bound inner core.

\section{Implications for Dark Matter Detection}

In direct detection experiments the differential interaction rate is sensitive to the fine grained density and the velocity distribution of dark matter particles on A.U. scales \cite{Copi1999,Copi2007}.  Substructures like microhalos can affect the interaction rate if they are abundant enough to have a substantial likelihood of existing in the solar neighbourhood and if their density is at least the same order of magnitude as the background dark matter density in this region,
$ \rho_{bg}\sim 10^{7} M_{\odot}$kpc$^{-3}$ (see for example \cite{Catena2009}).

Our results above suggest that none of these conditions are generally achieved. In Figure \ref{stream_density} we plot the stream densities of microhalos that crossed the stellar field for 0 Myr (red), 80 Myr (grey) and 160 Myr (blue), before orbiting in the galactic potential for 10 Gyr. The tidal streams of the initially unperturbed halo (red) have an average density of $\rho\sim 10^4 M_{\odot}$kpc$^{-3}$, which is already negligibly low compared to the background. Only the very tiny core still maintains its initial density of $\rho \approx 10^{11} M_{\odot}$kpc$^{-3}$. The initially disrupted microhalo (blue) has no more bound core. Its stream density is only  $\rho\sim 10^2-10^3 M_{\odot}$kpc$^{-3}$. 
The actual stream density of an average microhalo should therefore lie somewhere between the red and the blue line in Figure \ref{stream_density}.

Since the stream densities are far below the value of the local galactic density, only a surviving core existing in the region of the earth would any effect upon direct detection. However, only about half of the microhalos still have bound cores because of disk crossing, and tidal effects further reduce the mass of the cores to less than ten percent of their original value. We also note that any substructures orbiting primarily within the disk plane would be quickly destroyed by stellar encounters. Taking into account all these arguments leads to a chance of only about $0.0001\%$ to be in an overdense region today (for more detaile see \cite{Schneider2010}).

The streams of particles stripped from microhalos are coherent and long, thus it is appropriate to calculate their volume filling factor. Since the stream density is $\rho\sim 10^2-10^4 M_{\odot}$kpc$^{-3}$ we expect that our solar system is criss-crossed with $f_b \times (10^3-10^5)$ streams, where $f_b\approx0.1$ is the fraction of the local Galactic halo density that forms from substructures up to a solar mass. Larger substructures may be completely disrupted at the Sun's position in the Galaxy due to global disk shocking and tides \cite{dOnghia2010}. The velocity dispersion within an average stream due to heating by disk stars is $\sigma\sim10^{-2}$ km/s. Thus, the local density is determined by the superposition of a large number of independent streams, and the overall velocity distribution at the solar radius should be essentially Maxwellian, isotropic and smooth with nearly no spiky structure, as we would assume for a smooth halo model with no substructures. The signatures of streams could be only be detected experimentally with over several hundred events.

The case for indirect detection is somewhat different from that described above. In indirect detection experiments one tries to detect the annihilation products, such as gamma-rays, coming from the highest density dark matter regions, which is proportional to the square of the dark matter density times the volume of the region observed \cite{Lake1990,Bertone2005}.  Consider a volume containing on average one microhalo $V \approx 10^{-2}$ pc$^{-3}$. The luminosity due to the smooth background is therefore $L_{bg} \propto V \rho^2 = 10^{-6} M_{\odot}^2 $pc$^{-3} $, whereas the luminosity of a surviving microhalo core is $L_{mh} \propto V_{core} \rho_{core}^2 \approx 5 \times 10^{-7} M_{\odot}^2 pc^{-3}$. Here we have assumed a mean core density of $10^{10} M_{\odot} $pc$^{-3}$. Thus the net boost factor due to microhalos is about 1.5, and stays below the detection limits of the FERMI experiment \cite{Pieri2005}. However this number is highly uncertain since it depends on extrapolations of both the substructure mass function and also on the microhalo internal density structure.

Let us summarise: It has recently been shown that hierarchical clustering continues down to extremely small mass scales, so that most dark matter currently in the halo of our galaxy may have originated in microhalos with masses as small as $10^{-6} M_{\odot}$. These structures are very dense so that despite the various disruption processes, a couple of percent of the initial mass may still be within bound cores, while the rest is located in cold tidal streams. However, neither the surviving cores nor the tidal streams have a significant effect on direct or indirect detection. The limits obtained on dark matter from detection experiments under the conservative assumption of a smooth halo with nearly Maxwellian density distribution remain valid, with a prefactor depending only on the average local dark matter density.

\end{document}